\documentclass[letter]{aa}
\usepackage{graphicx}
\usepackage{txfonts}
\usepackage[colorlinks=true, citecolor=blue, linkcolor=blue, urlcolor=blue]{hyperref}

\begin{document}

    \title{The quasar main sequence of highly accreting NLSy1 galaxies}

    \titlerunning{The EV1 plane of highly accreting NLSy1 galaxies}

\author{Alberto Dom\'inguez\inst{1}\fnmsep\thanks{\email{alberto.d@ucm.es}}
        \and
       Suvendu Rakshit\inst{2}
       \and
       Vaidehi~S. Paliya\inst{3}
       \and
       D. J. Saikia\inst{4,5}
       \and
        C. S. Stalin\inst{6}
       }

\institute{$^1$IPARCOS and Department of EMFTEL, Universidad Complutense de Madrid, E-28040 Madrid, Spain \\
            $^2$Aryabhatta Research Institute of Observational Sciences, Nainital–263001, Uttarakhand, India\\
            $^3$Inter-University Centre for Astronomy and Astrophysics (IUCAA), SPPU Campus, 411007, Pune, India\\
            $^4$Fakult\"at f\"ur Physik, Universit\"at Bielefeld, Postfach 100131, D-33501 Bielefeld, Germany\\
            $^5$Assam Don Bosco University, Guwahati 781017, Assam, India\\
            $^6$Indian Institute of Astrophysics, Bengaluru, 560034, India
            }

   \date{Received February 24, 2026; accepted October 29, 2026}

\abstract
{The quasar main sequence, or Eigenvector 1 (EV1), describes the optical diversity of active galactic nuclei, with Narrow-Line Seyfert 1 (NLSy1) galaxies anchoring the high-accretion end. Recent discoveries of overly massive black holes in the early Universe highlight the need to study low-redshift, low-mass super-Eddington accretors as analogs for rapid black hole seed growth.}
{We aim to map a newly discovered, large population of 18,749 NLSy1 galaxies identified in the Dark Energy Spectroscopic Instrument Data Release 1 (DESI-DR1) onto the EV1 plane and to compare them with the SDSS-DR17 NLSy1 population analyzed in a homogeneous way.}
{We compare the spectral properties of the DESI-DR1 NLSy1 sample with the SDSS-DR17 NLSy1 catalog. We extract key parameters, including broad H$\beta$ full width at half maximum (FWHM) and Fe II strength ($R_{4570}$). To evaluate their accretion states, we derive single-epoch virial black hole masses using two approaches: an Fe II strength-dependent scaling relation, and a recently established Eddington ratio-dependent fundamental plane. We further assess how strongly the inferred accretion states depend on this choice by repeating the analysis with a conventional Fe~II-independent calibration.}
{The DESI-DR1 NLSy1 population occupies the extreme end of the EV1 parameter space, possessing strong Fe II emission (median $R_{4570} = 0.96\pm0.35$), similar to the SDSS sample ($0.96\pm0.36$). Under both calibrations the two samples harbor low-mass black holes (median $\log M_{\text{BH}}/M_{\odot} \approx 6.81-6.85$) and $\sim50\%$ formally exceed the Eddington limit ($\log \lambda_{\rm Edd} > 0$); this fraction drops to $\sim8-9\%$ with the Fe~II-independent calibration, and is therefore strongly calibration dependent.}
{The unprecedented sensitivity of DESI has unveiled a large population of low-mass, highly accreting NLSy1 galaxies. Under the high-accretion interpretation, the inner disk is expected to become geometrically thick and to shield the outer broad-line region, naturally producing the observed strong Fe~II emission. This sample provides a rich statistical dataset of low-redshift candidate super-Eddington accretors, serving as laboratories for understanding early-Universe black hole growth.}


   \maketitle
\nolinenumbers

\section{Introduction}\label{sec:intro}

The optical diversity of active galactic nuclei (AGN) is largely driven by a parameter space known as the quasar main sequence, or Eigenvector 1 (EV1) \citep[e.g.,][]{Boroson1992,Sulentic2000,Boroson2002,Shen2014}. It is defined by an anti-correlation between the optical Fe~II strength, parameterized by the flux ratio $R_{4570} = \text{Fe~II} / \text{H}\beta$, and the [O~III] $\lambda 5007$ line, alongside the full width at half maximum (FWHM) of the broad H$\beta$ component. EV1 is widely interpreted as being governed by the Eddington ratio ($\lambda_{\rm Edd} = L_{\rm bol}/L_{\rm Edd}$), with secondary dependencies on black hole mass ($M_{\rm BH}$) and the observer's viewing angle \citep[e.g.,][]{Marziani2001, Shen2014}, and its detailed morphology is commonly described through the Population~A/B classification \citep[e.g.,][]{Sulentic2000, Zamfir2010}.

Narrow-Line Seyfert 1 (NLSy1) galaxies anchor the high-accretion end of EV1. Classified by a broad H$\beta$ FWHM $\leq 2000 \text{ km s}^{-1}$ and an [O~III]/H$\beta$ ratio $< 3$ \citep[e.g.,][]{Osterbrock1985, Goodrich1989}, they frequently show strong Fe~II complexes and rapid X-ray variability \citep[e.g.,][]{Boller1996, Leighly1999a}, signatures of relatively low-mass black holes ($M_{\text{BH}} \sim 10^{6-8} M_{\odot}$) accreting at or near the Eddington limit \citep[e.g.,][]{Mathur2000, Grupe2004}.

As of now, large-scale statistical studies of EV1 have heavily relied on optical spectra from the Sloan Digital Sky Survey (SDSS) \citep[e.g.,][]{Shen2011, Rakshit2017, Paliya2024}. While SDSS has been instrumental in characterizing the bright AGN population, its flux limits may introduce an observational bias against the fainter, low-mass AGN that populate the faint end of EV1.

The necessity of characterizing these efficient accretors has been further amplified by recent discoveries from the James Webb Space Telescope (JWST). Observations of overly massive black holes at redshifts $z > 6$ challenge traditional formation models \citep[e.g.,][]{Maiolino2023, Bogdan2023}, suggesting that early black hole seeds may have undergone sustained phases of super-Eddington accretion to reach such masses within the first billion years of the Universe \citep[e.g.,][]{Volonteri2021, Pacucci2022}. Therefore, finding and studying low-mass, super-Eddington AGN in the low-redshift Universe is crucial, as they may serve as accessible analogs for investigating the physics of rapid black hole growth.

The recent Data Release 1 (DR1) of the Dark Energy Spectroscopic Instrument \citep[DESI;][]{DESI2025} provides a unique opportunity to overcome previous limitations. With its significantly deeper sensitivity and broader sky coverage compared to SDSS, the DESI survey allows for a robust exploration of the low-luminosity AGN regime. Recently, we performed a detailed spectral decomposition of over 71,000 optical spectra from DESI-DR1, identifying 18,749 NLSy1 galaxies for the first time \citep{Paliya2026}. In this paper, we aim to map this newly discovered, faint NLSy1 population onto the EV1 plane. By doing so, we will investigate their accretion properties and determine whether these low-mass rapidly accreting sources represent an extended tail of the quasar population.

\section{Data and sample selection}\label{sec:data}

\subsection{The DESI-DR1 and SDSS-DR17 NLSy1 catalogs}\label{sec:catalogs}

To investigate the EV1 properties, we utilized the recently compiled catalog of 18,749 NLSy1 galaxies identified in the Dark Energy Spectroscopic Instrument Data Release 1 (DESI-DR1). This sample was constructed by performing a systematic spectroscopic decomposition of over 71,000 optical spectra of AGN located at $z \le 0.9$. The continuum-subtracted spectra were modeled using a combination of a Lorentzian profile for the broad H$\beta$ component and Gaussian profiles for the narrow emission lines \citep{Paliya2026}. The objects in the DESI-DR1 sample were selected based on the traditional NLSy1 optical classification criteria: (1) The full width at half maximum (FWHM) of the broad H$\beta$ component is less than $2000~\text{km~s}^{-1}$, accounting for measurement uncertainties. (2) The flux ratio of the narrow $[\text{O~III}]~\lambda5007$ emission line to the total H$\beta$ emission line is $< 3$.

For the comparison study, we used the catalog of 22,656 NLSy1 galaxies identified in SDSS-DR17 \citep[][]{Paliya2024}. Both catalogs adopt identical selection criteria and the same broad H$\beta$ profile, although their Fe~II templates differ (Sect.~2.2). To avoid possible biases due to different observing strategies of DESI and SDSS surveys, we matched each SDSS NLSy1 to one DESI NLSy1 in redshift ($\Delta z<0.05$), rest-frame monochromatic continuum luminosity at $5100~\text{\AA}$ ($\Delta \log \lambda L_{\lambda,5100} < 0.1$~dex), and spectral signal-to-noise ratio ($\Delta S/N<0.1$), minimizing $(\Delta z/0.05)^2+(\Delta \log \lambda L_{\lambda,5100}/0.1)^2+(\Delta S/N/0.1)^2$ if several qualified, leading to 11,184 pairs.

\subsection{Key spectral parameters}\label{sec:params}

To map these populations onto the EV1 plane and estimate their accretion properties, we extracted several key parameters directly from the spectral fitting catalogs. The broad H$\beta$ FWHM corresponds to the velocity width of the broad H$\beta$ line ($v_{\text{BLR}}$), which acts as a proxy for the virial velocity of the gas in the broad-line region (BLR). The Fe~II strength ($R_{4570}$) quantifies the relative strength of the optical Fe~II emission and is defined as the ratio of the integrated Fe~II pseudo-continuum flux within the wavelength range of $4434~\text{\AA}$ to $4684~\text{\AA}$ to the flux of the broad H$\beta$ component. The 5100 \AA~continuum luminosity was used to estimate the bolometric luminosity ($L_{\text{bol}}$) following the bolometric correction factor of 9.26 proposed by \citet[][]{Richards2006}.

The SDSS-DR17 catalog used the Fe~II template of \citet[][]{VCV04}, whereas DESI-DR1 used that of \citet[][]{Boroson1992}. To avoid systematics from this difference, we re-analyzed the SDSS spectra with the pipeline used for the DESI-DR1 catalog.
This re-analysis does not establish which Fe~II template is preferable; however, it ensures uniformity.
Moreover, 9434 SDSS-NLSy1 galaxies were also observed by DESI; we analyzed their DESI spectra to test directly for systematic offsets in $R_{4570}$ and FWHM (Appendix~A).
Throughout this work, values quoted after a median are median absolute deviations, characterizing the width of the distribution; the uncertainties on the medians themselves are below $0.01$ for $R_{4570}$, $0.01$~dex for logarithmic quantities, and $10~\text{km~s}^{-1}$ for FWHM, for samples of this size. These derived parameters provide the necessary foundation to evaluate the distribution of both samples across the $R_{4570}$ versus H$\beta$ FWHM parameter space and to derive their single-epoch $M_{\rm BH}$ and $\lambda_{\rm Edd}$ parameters.

\section{Black hole mass and Eddington ratio}\label{sec:mass}
To accurately place the DESI-DR1 and SDSS-DR17 NLSy1 samples within the context of the EV1 parameter space, we derived $M_{\rm BH}$ and $\lambda_{\rm Edd}$ values. In highly accreting AGN, the H$\beta$ lag is systematically shorter than predicted by the canonical $R_{\text{BLR}}$--$L_{5100}$ relation calibrated on sub-Eddington sources \citep[by factors of $\sim2-6$;][]{Du2015, Du2016}, so that applying the canonical relation overestimates $R_{\text{BLR}}$ and hence $M_{\text{BH}}$ for this population \citep[e.g., by up to a factor of five for super-Eddington DESI quasars;][]{Pan2025}. We therefore use two scaling relations designed to correct for high-Eddington biases. First, we adopt the $M_{\text{BH}}$ values of the primary catalogs \citep{Paliya2026}, derived from the \citet{Du2019} relation, which includes a correction for the Fe~II strength ($R_{4570}$). Second, we recompute $M_{\text{BH}}$ for both samples with the Eddington ratio-dependent fundamental plane of \citet{Woo2026}, based on $L_{5100}$ and the broad H$\beta$ FWHM.

\begin{figure*}
\centering
\includegraphics[scale=0.3]{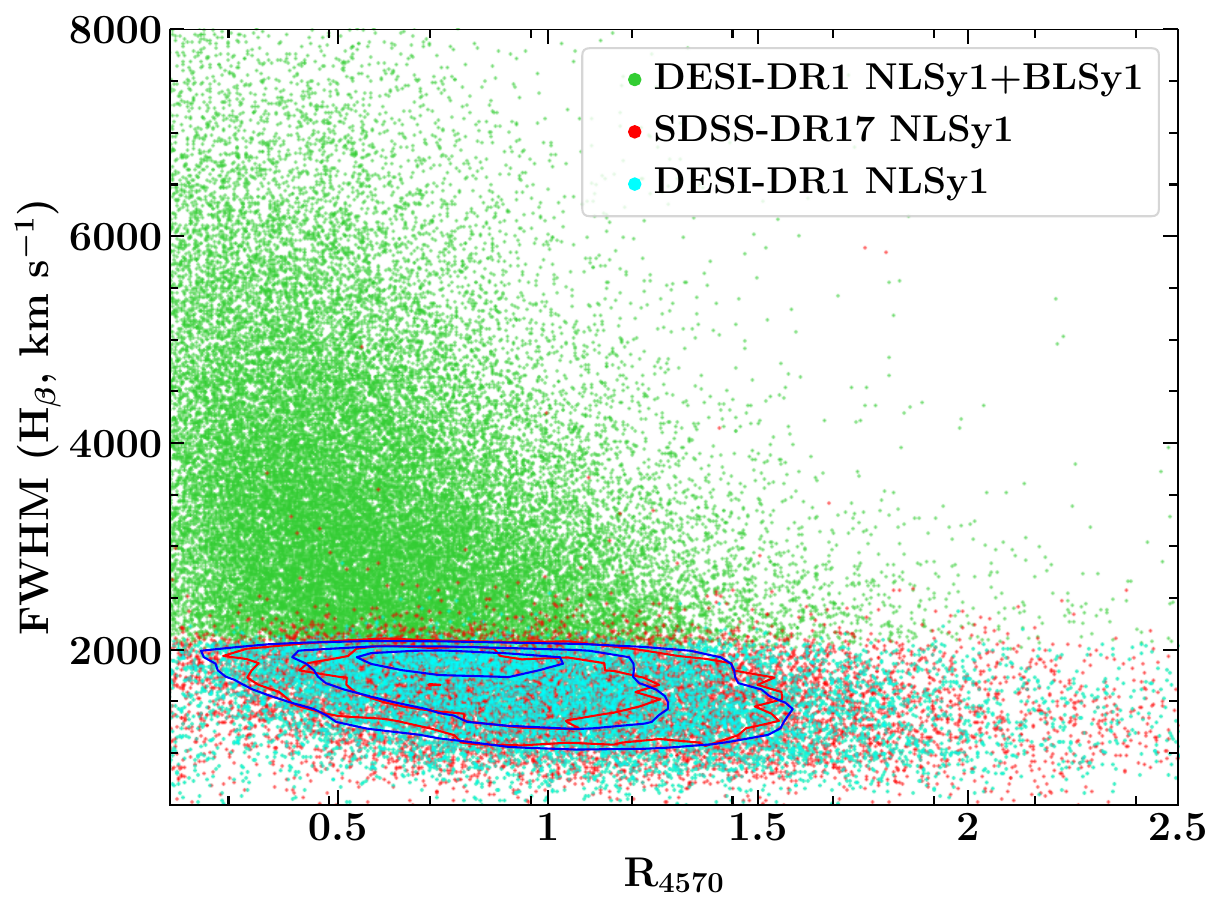}
\includegraphics[scale=0.3]{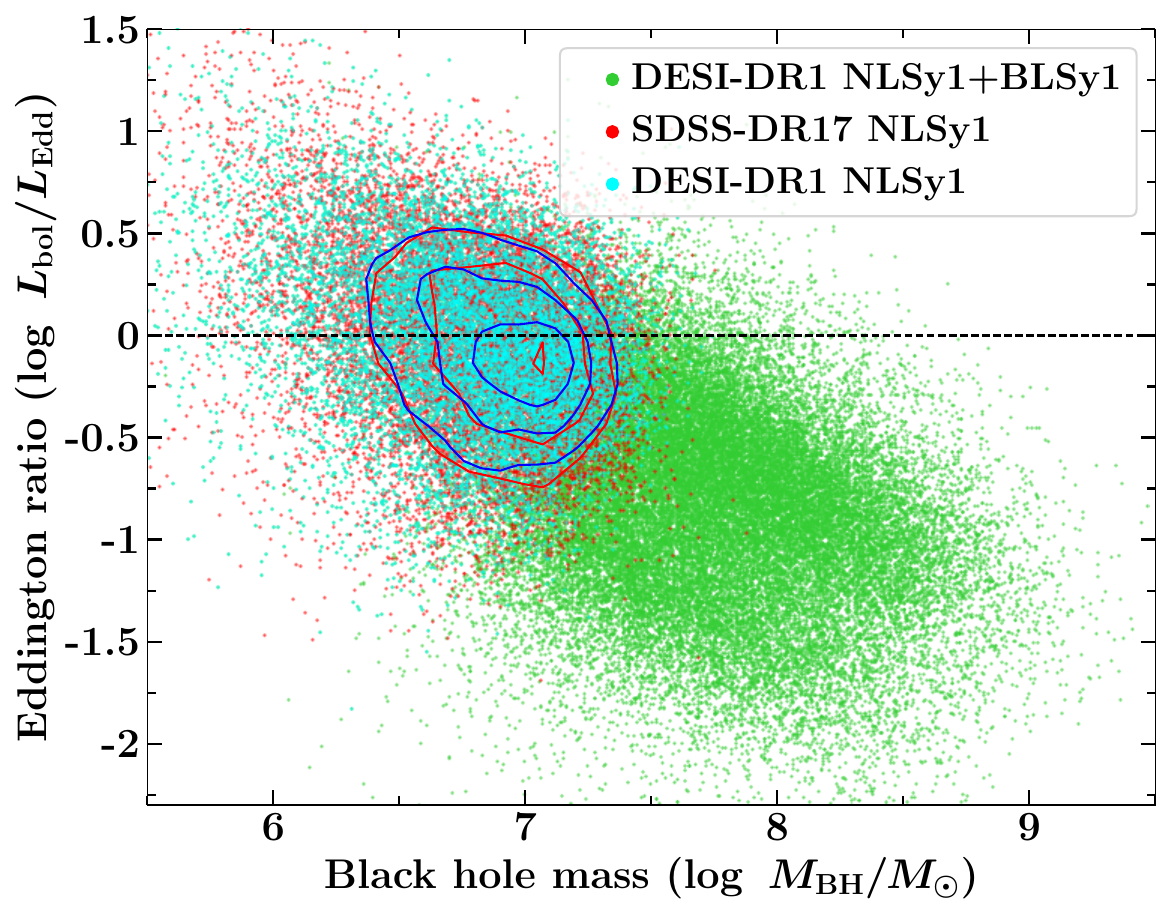}
\caption{(Left) The EV1 diagram displaying Fe~II strength ($R_{4570}$) versus broad H$\beta$ FWHM for the matched samples of DESI and SDSS NLSy1 galaxies. The DESI-DR1 parent AGN sample (BLSy1 + NLSy1) is shown in the background (green) to illustrate how NLSy1 galaxies populate an extended tail of EV1. Smoothed contours trace 40\%, 65\%, and 85\% of the maximum kernel density of each NLSy1 sample; they are iso-density levels and do not enclose these fractions of the objects. (Right) Eddington ratio versus $M_{\rm BH}$ using the \citet{Du2019} calibration.
}
\label{Figure1}
\end{figure*}

\section{Results}\label{sec:results}

\subsection{The Eigenvector 1 plane}\label{sec:ev1}

Figure~\ref{Figure1} displays the distributions of the DESI and SDSS NLSy1 samples in the $R_{4570}$ versus broad H$\beta$ FWHM plane.
As can be seen, both SDSS and DESI NLSy1 samples occupy the lower quadrant of the EV1 plane, bounded by the $2000~\text{km~s}^{-1}$ FWHM limit. Quantitatively, the median value of the Fe~II strength ($R_{4570}$) for the DESI-NLSy1 population is $0.96 \pm 0.35$, which is similar to the median of $0.96 \pm 0.36$ observed for the SDSS-NLSy1 sources.
These results indicate that, within the considered diagnostics,
the DESI and SDSS samples are consistent with tracing the same population of extreme EV1 objects within the general quasar population.

\subsection{The high-accretion regime}\label{sec:highacc}

The position of the DESI sample on the EV1 plane, characterized by narrow broad lines and strong Fe~II emission, points toward high accretion rates. We compared the accretion properties of the matched samples by evaluating their $\lambda_{\rm Edd}$ as a function of their $M_{\rm BH}$ values. Figure~\ref{Figure1} illustrates this parameter space. Our analysis reveals that the DESI sample harbors low mass black holes with a median value of $\log(M_{\text{BH}}/M_{\odot}) = 6.81 \pm 0.26$, whereas the SDSS sample shows a similar median mass of $\log(M_{\text{BH}}/M_{\odot}) = 6.83 \pm 0.26$.
Using the measured $L_{\text{bol}}$ values, we derived the Eddington ratio and found a large fraction of both samples to lie formally on or above the Eddington limit ($\log \lambda_{\text{Edd}} > 0$) in the $\lambda_{\text{Edd}}$--$M_{\text{BH}}$ plane. When utilizing the \citet{Du2019} calibration, the logarithmic median $\lambda_{\rm Edd}$ is $0.01 \pm 0.29$ for the DESI sample, compared to $0.00 \pm 0.32$ for SDSS, with $\sim$50\% of the DESI and SDSS samples formally crossing the Eddington limit. Applying the \citet{Woo2026}
calibration, we found similar results with $\log(M_{\text{BH}}/M_{\odot}) = 6.83\pm0.28$ and $6.85\pm0.28$ for DESI and SDSS samples, respectively. The estimated logarithmic $\lambda_{\rm Edd}$ values are $0.00 \pm 0.24$ and $-0.02\pm0.25$, respectively. Approximately, 50\% and 48\% of the DESI and SDSS NLSy1s formally exceed the Eddington limit.

Since the estimated super-Eddington fraction could be sensitive to the intrinsic scatter of single-epoch mass relations, measurement errors and the bolometric correction, we performed a Monte Carlo test by propagating the cataloged uncertainties in FWHM and $\lambda L_{\lambda,5100}$, the intrinsic scatter in the mass estimators, and the uncertainty in the bolometric correction. The resulting $16$th--$84$th percentile ranges of the super-Eddington fractions for DESI and SDSS NLSy1s are $43-59$\% and $43-57$\% for the \citet{Du2019} calibration. Using the calibrations of \citet{Woo2026}, we obtained super-Eddington fractions of $42-59$\% and $41-57$\% for the DESI and SDSS populations, respectively  (see Appendix B for details).

Under calibrations correcting for the shortened H$\beta$ lags of high-accretion sources, the EV1 location of these galaxies is thus accompanied by Eddington ratios close to unity. The fraction formally above the Eddington limit is, however, strongly calibration dependent, falling to $\sim8-9\%$ with the Fe~II-independent relation of \citet{Vestergaard2006} (Sect.~\ref{sec:caveat}). The robust result is observational, namely that these sources occupy the strong-Fe~II, narrow-H$\beta$ end of EV1, while their identification as super-Eddington accretors remains a calibration-dependent inference. They are nevertheless potential low-redshift analogs of the overly massive high-redshift black holes discovered by JWST \citep[e.g.,][]{Maiolino2023, Bogdan2023}.

\section{Discussion}\label{sec:discussion}

\subsection{Physical drivers of the EV1 location}\label{sec:drivers}

The location of the NLSy1 population toward the bottom-right of the EV1 plane probes the physics of high accretion \citep{Shen2014}. The primary physical driver of EV1 is $\lambda_{\rm Edd}$, for which the optical Fe~II/H$\beta$ ratio serves as a useful observational proxy \citep[e.g.,][]{Gaskell2022}. As the Eddington ratio approaches or exceeds unity, the inner accretion disk geometry is expected to expand vertically, transitioning into a slim disk configuration \citep{Abramowicz1988}.

This geometrically thick disk can shield the outer BLR from the intense, ionizing extreme-ultraviolet continuum emitted by the innermost regions \citep[the self-shadowing effect described in detail by][]{Wang2014}. This shielding creates an optimal, dense, partially neutral environment that is highly conducive to the production of Fe~II emission \citep{Ferland2009}. Simultaneously, the radiation pressure associated with such high accretion rates may push the BLR gas outward, lowering the Keplerian velocity and resulting in the narrow broad H$\beta$ emission lines ($< 2000~\text{km~s}^{-1}$) that define the NLSy1 class \citep{Osterbrock1985}. Therefore, the low $M_{\rm BH}$ values of NLSy1 sources ($\sim 10^{6.8} M_{\odot}$) drive them to high $\lambda_{\rm Edd}$ at comparable continuum luminosities. Such a high-accretion state may trigger the slim disk shielding, producing the extreme $R_{4570}$ values observed. If these sources are accreting in the super-Eddington regime, the associated radiation pressure may also facilitate relativistic jet formation \citep{Ojha2026}, as seen in a subset of radio-loud NLSy1s, including those showing large-scale double-lobed structures \citep{Umayal2025}, potentially enhancing feedback on host galaxy scales \citep{Salome2023}.

We caution that $L_{\text{bol}}/L_{\text{Edd}} > 1$ may not be equivalent to a super-Eddington mass inflow rate: in the slim-disk regime photon trapping causes the emergent luminosity to saturate roughly logarithmically with $\dot{M}$ \citep{Abramowicz1988, Wang2014}. Strong optical Fe~II emission has also been linked to rapid, recent nuclear star formation and iron enrichment \citep{Collin2000}; we stress, however, that $R_{4570}$ is not a direct abundance diagnostic, since it also depends on gas density, column density, the shape of the ionizing continuum, micro-turbulence, covering factor and radiative-transfer conditions \citep[e.g.,][]{Ferland2009, Gaskell2022}.

\subsection{Implications for black hole seed growth}\label{sec:seeds}

JWST has revealed overly massive black holes at redshifts $z > 6$ that challenge standard formation models \citep[e.g.,][]{Maiolino2023, Bogdan2023}; to grow to such masses within the first billion years, these seeds might have undergone extended periods of super-Eddington accretion \citep[e.g.,][]{Volonteri2021, Pacucci2022}. The highly accreting low-mass black holes of the NLSy1 population represent a potential low-redshift laboratory for studying this growth phase, through their spectral energy distributions, outflow signatures and metallicities.

\subsection{Caveats and systematics}\label{sec:caveat}

Because the physical interpretation of the DESI and SDSS NLSy1 samples relies heavily on their high inferred $\lambda_{\rm Edd}$ values, we assessed how strongly they depend on the adopted mass calibration. To this end, we carried out a simple test using a conventional Fe~II-independent H$\beta$ single-epoch relation of \citet[][]{Vestergaard2006}. This exercise allowed us to study the dependence of the numerical results on the adopted high accretion corrections proposed by \citet[][]{Du2019} and \citet[][]{Woo2026}. The calculated logarithmic $M_{\rm BH}$ values (in $M_{\odot}$) for the DESI and SDSS populations are $7.24\pm0.24$ and $7.26\pm0.25$, respectively. The corresponding $\lambda_{\rm Edd}$ values are $-0.41\pm0.21$ and $-0.43\pm0.21$, respectively, and only 8\% and 9\% of the DESI and SDSS NLSy1 populations formally exceed the Eddington limit. Therefore, the fraction formally exceeding the Eddington limit, as well as the inferred black hole masses and Eddington ratios, depends strongly on the adopted $M_{\rm BH}$ calibration (Appendix C). However, additional evidence consistent with high accretion rates is provided by the strong Fe~II complexes observed in these sources, as expected from recent CLOUDY and FRADO calculations \citep[see also,][]{Panda2019,Naddaf2025}. We emphasize that this is supporting, qualitative evidence rather than an independent measurement of the accretion rate: $R_{4570}$ enters the \citet{Du2019} calibration explicitly, and is at the same time one of the observables that defines the EV1 location being interpreted.
Finally, relations between $R_{\rm BLR}$, virial mass and inferred Eddington ratio depend on the calibration sample, mass prescription, bolometric correction, inclination distribution and data quality; they may trace broad population trends but should not be treated as universal quantitative calibrations \citep[e.g.,][]{Naddaf2025}.

A further caveat concerns the sample definition. NLSy1 galaxies are selected by requiring a broad H$\beta$ FWHM below $2000~\text{km~s}^{-1}$, so part of their location along the FWHM axis follows from the selection itself; and since FWHM enters the virial estimators approximately quadratically, this also favors lower inferred $M_{\rm BH}$ and, at fixed luminosity, higher inferred $\lambda_{\rm Edd}$. The narrow-line location, the low masses and the high Eddington ratios are therefore not statistically independent findings. What remains genuinely informative is the Fe~II behavior, which is not used in the selection, the distribution of the population within the selected region, and the fact that two samples selected identically and analyzed homogeneously are closely similar.

Alternative models propose that NLSy1 galaxies are standard broad-line Seyfert 1 galaxies viewed nearly face-on \citep[e.g.,][]{Decarli2008}, so that projection effects narrow the observed H$\beta$ FWHM and bias the virial masses downward. Orientation likely contributes to the FWHM scatter, but does not readily explain the enhanced Fe~II emission seen in both samples.

\section{Summary and conclusions}\label{sec:conclusions}

We have projected the sample of 18,749 DESI-DR1 NLSy1 galaxies \citep{Paliya2026} onto the EV1 plane and compared it with the SDSS-DR17 NLSy1 population \citep{Paliya2024}, reaching the following conclusions: (1) The DESI and SDSS surveys have uncovered an intriguing tail of EV1. Both DESI and SDSS NLSy1 samples, matched in redshift, $\lambda L_{\lambda,5100}$, and spectral signal-to-noise ratio, show strong Fe~II emission (median $R_{4570} = 0.96\pm0.35$ and $0.96\pm0.36$, respectively). (2) The unique position of NLSy1 galaxies on the EV1 plane is accompanied by low inferred black hole masses, which partly follow from the FWHM-based selection (Sect.~\ref{sec:caveat}). The median single-epoch virial masses for the DESI and SDSS samples are $\log(M_{\text{BH}}/M_{\odot}) = 6.81 \pm 0.26$ and $6.83 \pm 0.26$, respectively. (3) Coupling their lower $M_{\rm BH}$ with their continuum luminosities reveals that a substantial fraction of the NLSy1 population ($\sim50\%$) formally lies at or above the Eddington limit under both the \citet{Du2019} and \citet{Woo2026} calibrations. This fraction is, however, strongly dependent on the adopted mass calibration, falling to $\sim8-9\%$ with the Fe~II-independent relation of \citet{Vestergaard2006}. These objects are therefore best described as highly accreting NLSy1 galaxies and as candidate super-Eddington accretors, rather than as a population with a securely measured super-Eddington fraction.

Ultimately, this EV1 sample provides a rich statistical dataset of low-redshift candidate super-Eddington accretors. Future multi-wavelength investigations, including reverberation mapping to calibrate their virial masses and high-resolution host galaxy imaging, will be pivotal in unraveling the physics governing the extreme faint end of the AGN population.

\section*{Data availability}
Catalogs and spectral parameters are available at \url{https://www.ucm.es/blazars/seyfert} and Zenodo (\url{https://doi.org/10.5281/zenodo.20484681}). DESI-DR1 and SDSS-DR17 data are accessible via their respective survey archives.

\begin{acknowledgements}
We thank the anonymous referee for a constructive report.
AD acknowledges support from MCIN/AEI under grant PID2025-171816NB-I00 (MCIN/AEI/10.13039/501100011033/FEDER, EU).
\end{acknowledgements}

\begin{appendix}
\nolinenumbers

\section{Comparison of the NLSy1 galaxies observed both with DESI and SDSS}
\label{app:1}

\begin{figure}[!h]
\centering
\includegraphics[width=\linewidth]{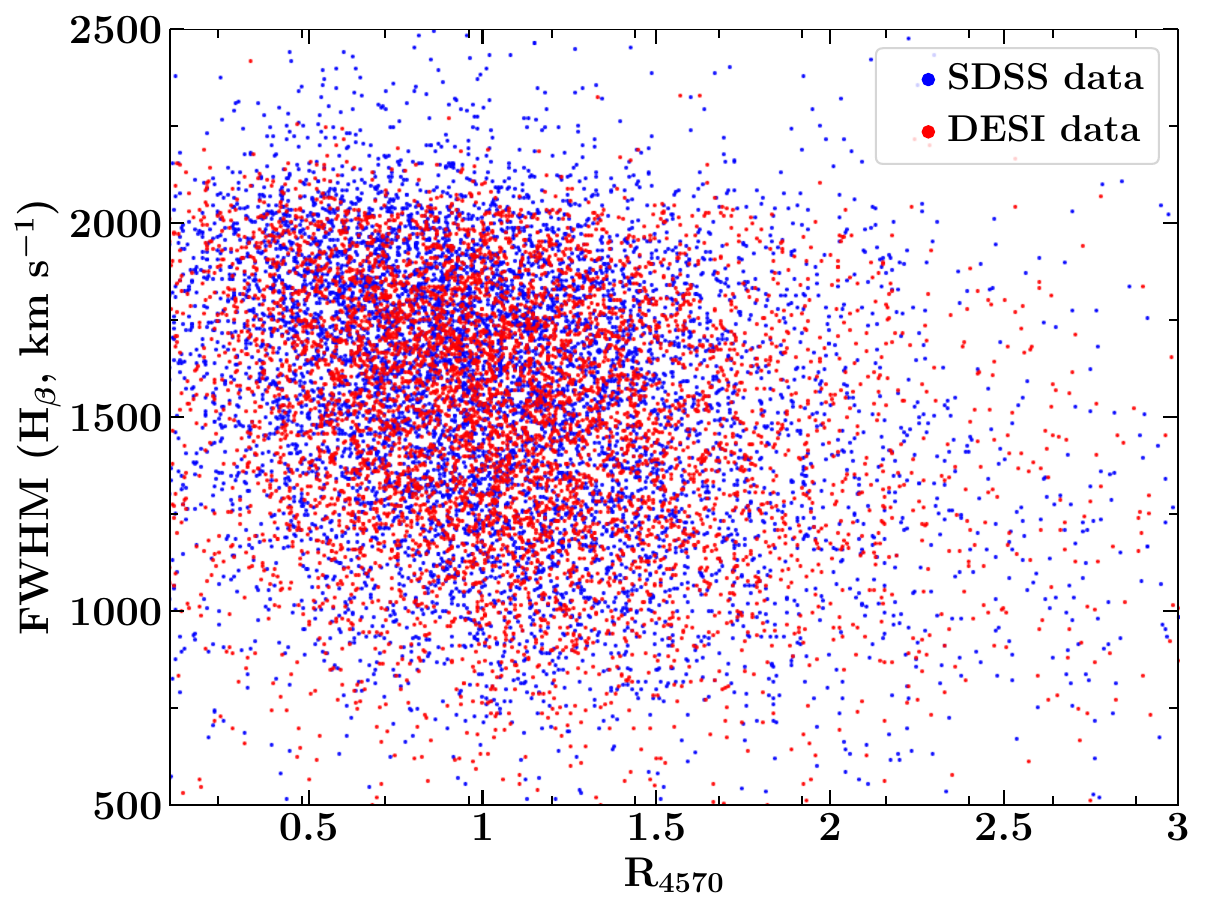}
\caption{The EV1 diagram displaying $R_{4570}$ versus broad H$\beta$ FWHM for the 9434 NLSy1 galaxies observed with both DESI (red) and SDSS (blue).}
\label{Figure1_app}
\end{figure}

There are 9434 SDSS-DR17 NLSy1 galaxies whose optical spectra were also taken with the DESI survey. To get an idea of the possible offset and systematics in the measurement of the FWHM of the broad H$\beta$ line and $R_{4570}$, we analyzed the DESI spectra of these objects with the same pipeline used to produce the DESI-DR1 NLSy1 catalog. In Figure~\ref{Figure1_app}, we show the position of these objects on the EV1 plane. As can be seen, there is no practically important difference in the measured spectral properties, which is similar to that found for the matched sample of DESI-DR1 and SDSS-DR17 NLSy1 sources (Figure~\ref{Figure1}). In particular, the median broad H$\beta$ FWHM and $R_{4570}$ offsets between the SDSS and DESI spectra are $60$~km~s$^{-1}$ and $0.05$, with median absolute deviations of $361$~km~s$^{-1}$ and $0.49$, respectively. Both offsets are negligible compared with the widths of the distributions, indicating that residual instrumental differences between the two surveys are unlikely to dominate the comparison.

\section{Monte Carlo uncertainties on the super-Eddington fractions}
\label{app:2}

Because the median Eddington ratios of both samples lie close to $\log \lambda_{\text{Edd}} = 0$, the fraction of objects formally above the Eddington limit is sensitive to the errors on the input quantities. We therefore generated 20000 Monte Carlo realizations of both catalogs, perturbing in each realization (i) the measured FWHM(H$\beta$) and $\lambda L_{\lambda,5100}$ by their cataloged uncertainties; (ii) the mass estimator by its intrinsic scatter, taken to be $0.3$~dex for the relations of \citet{Du2019} and \citet{Woo2026}; and (iii) the bolometric correction by $0.1$~dex. The cataloged FWHM and $\lambda L_{\lambda,5100}$ uncertainties were drawn independently per object, the mass-estimator scatter was added in quadrature, and the bolometric correction was drawn once per realization so that it acts coherently on the whole sample. In detail, the cataloged uncertainties were propagated to $\log \lambda_{\rm Edd}$ through the partial derivatives of each mass relation, combined in quadrature with the $0.3$~dex scatter, and a Gaussian deviate with this dispersion was drawn independently for every object in every realization. The value of $0.3$~dex is adopted as a representative intrinsic scatter of reverberation-calibrated single-epoch H$\beta$ estimators \citep[$0.3$--$0.4$~dex;][]{Shen2008}, and the same value is applied to both relations for uniformity. The $0.1$~dex on the bolometric correction is the typical dispersion of the correction around the universal-SED value \citep{Shen2008}. We apply it coherently to the whole sample as a conservative choice, since the fraction above the limit is most sensitive to a coherent shift and NLSy1s may deviate systematically from the mean SED.

For the DESI and SDSS populations, the resulting $16$th--$84$th percentile ranges of the super-Eddington fractions are $43-59$\% and $43-57$\%, respectively, for the \citet{Du2019} calibration. Considering \citet{Woo2026}, we estimated the super-Eddington fraction to be $42-59$\% and $41-57$\%. Each fraction is thus uncertain by $\sim\pm8$ percentage points, while the two samples track each other to within $\sim1$ percentage point in every realization, because the dominant systematic acts coherently on both populations.

\section{Sensitivity to the adopted mass calibration}
\label{app:3}

As a sensitivity test we recomputed the masses of both samples with the conventional, Fe~II-independent single-epoch H$\beta$ relation of \citet{Vestergaard2006},
\begin{equation}
\log \left( \frac{M_{\text{BH}}}{M_{\odot}} \right) = 6.91 + 2 \log \left( \frac{\mathrm{FWHM}}{1000~\mathrm{km~s}^{-1}} \right) + 0.50 \log \left( \frac{\lambda L_{\lambda,5100}}{10^{44}~\mathrm{erg~s}^{-1}} \right).
\label{eq:vp06}
\end{equation}
This prescription does not correct for the shortened lags of high-accretion sources and therefore returns systematically larger masses and lower Eddington ratios; it is not intended to provide more accurate masses, but to show how strongly the numerical results depend on the adopted high-accretion correction. As discussed in Section~\ref{sec:caveat}, the estimated $M_{\rm BH}$ values are substantially larger than those obtained using \citet[][]{Du2019} and \citet[][]{Woo2026} calibrations.

The virial factor $f$ introduces a further, coherent systematic. For a flattened BLR, $f$ scales approximately as $(a^{2} + \sin^{2} i)^{-1}$ for a disk of thickness parameter $a$ observed at inclination $i$ \citep[e.g.,][]{McLure2004}. Varying $f$ over the wide bracket spanned by $i = 10^{\circ}-40^{\circ}$ (with $a = 0.2$) shifts $\log M_{\text{BH}}$ by $0.81$~dex, and $\log \lambda_{\text{Edd}}$ by the same amount with opposite sign. Because NLSy1s selected in the same way from both surveys are expected to share a similar inclination distribution, this systematic moves the two populations coherently and leaves their comparison unchanged. It does, however, affect the absolute super-Eddington fraction of the considered samples, illustrating that the individual fractions carry no more weight than the virial factor assumed to derive them.

\end{appendix}

\end{document}